\documentstyle[psfig]{europhys}

\def\etal{{\hbox{{\tenit\ et al.\/}\tenrm :\ }}}

\def\And{{\rm and\ }}

\def\stars{\bigskip\centerline{***}\medskip}

\newif\ifboo \boofalse

\def\Review#1{\boofalse{\it #1},}
\def\Name#1{{\sc #1},}
\def\Vol#1{\ifboo Vol. {\bf #1}\else{\bf #1}\fi}
\def\Year#1{\ifboo #1\else(#1)\fi}
\def\Book#1{\bootrue{\it #1},}
\def\Page#1{\ifboo {\rm p. #1}\else{\rm #1}\fi}

\begin{document}

\euro{}{}{}{}
\Date{}
\shorttitle{M. FORNARI \etal FLOATING BONDS AND GAP STATES IN a-SI ETC.}
\title{Floating bonds  and  gap states 
in a-Si and a-Si:H \\from first principles calculations}

\author{M. Fornari\inst{1}\footnote{Present address:
Naval Research Laboratory - Code 6391, Washington DC  20375-5345; 
e-mail: {\tt fornari@dave.nrl.navy.mil}}, 
M. Peressi\inst{1}, S. de Gironcoli\inst{2},  
and A. Baldereschi\inst{1,3} }
\institute{
     \inst{1}  Istituto Nazionale di Fisica della Materia (INFM) and 
Dipartimento di Fisica Teorica,\\ Universit\`a di Trieste, Strada
Costiera 11, I-34014 Trieste, Italy\\
\inst{2} Istituto Nazionale di Fisica della Materia (INFM) and  
Scuola Internazionale\\ Superiore di Studi Avanzati (SISSA), via Beirut 2-4,
 I-34014 Trieste, Italy\\
\inst{3}
Institut de Physique Appliqu\'ee, Ecole Polytechnique            
F\'ed\'eral de Lausanne, \\ PHB-Ecublens, CH-1015 Lausanne,
Switzerland}
 \rec{April 1999}{}
\pacs{
\Pacs{71}{23.$-$k}{Electronic structure of disordered solids}
\Pacs{71}{23.An}{Theories and models; localized states}
\Pacs{71}{23.Cq}{Amorphous semiconductors, metallic glasses, glasses}
}
\maketitle
\begin{abstract}
We  study in detail by means of ab-initio pseudopotential calculations
the electronic structure of 
five-fold coordinated ($T_5$) defects  in a-Si  and a-Si:H, also during
their formation and their evolution upon hydrogenation.
The atom-projected  densities of states (DOS) and an accurate 
analysis of the valence charge distribution clearly indicate 
the fundamental contribution of $T_5$ defects
in originating gap states through their nearest neighbors. 
The interaction with hydrogen can reduce 
the DOS in the gap annihilating $T_5$ defects.
\end{abstract}

The atomic origin of the midgap energy levels
in  a-Si has been  commonly ascribed  to
three-fold ($T_3$)  coordinated  atoms, 
and the observed reduction
upon hydrogenation has been explained with the passivation of the
``dangling bonds'' by H \cite{HinSi:generale}. This picture 
has been proposed in analogy with the well known mechanism  occurring in 
undercoordinated configurations such as 
 at surfaces and vacancies in crystalline silicon (c-Si)
\cite{Pickett81}. 

The dangling bond picture for a-Si has been widely  supported
 by many electronic structure calculations
\cite{fedders88,bisw89,hm93,InHoLee,tuttle96}, 
although it has been recognized 
that gap states can be induced also by other 
coordination defects \cite{bisw89,tuttle96}, e.g. five-fold ($T_5$)
coordinated ``floating bonds''   and {\em anomalous} four-fold ($T_4$) 
coordinated atoms.   

We focus our attention here on the role of $T_5$ defects.
Their importance in a-Si has been clearly stated  by 
Pantelides \cite{Pantelides,Pantelides87} and 
Kelires and Tersoff \cite{KT88} a douzen of years ago.
Pantelides \cite{Pantelides,Pantelides87} suggested 
that the presence and the role of $T_5$ defects must be seriously reconsidered 
in order to explain some theoretical and experimental 
data (effective electron correlation \cite{U}, hyperfine structure from 
electron paramagnetic resonance data \cite{Bigelsen},
relationship between intensity of paramagnetic signal and
density of Si-H bonds \cite{densitySiHbonds}) 
that otherwise would remain unexplained
in the common picture involving only  dangling bonds, and he 
gave some arguments 
suggesting that $T_5$ defects could be predominant. 
He  argued that $T_3$ and $T_5$ are conjugated defects,
since a bond elongation can transform a $T_4 + T_5$
structure into a $T_3+T_4$ one \cite{Pantelides};
furthermore, he proposed a mechanism for H diffusion based on
floating-bond switching and  annihilation/formation of $T_5$'s through 
interaction with H \cite{Pantelides87}.
The  empirical simulation by 
Kelires and Tersoff \cite{KT88} has shown that $T_5$
atoms have lower energy than $T_3$ atoms, and therefore should be favoured 
in general. Also some {\it ab-initio}
molecular dynamics simulations of a-Si structures indicate a
predominance of $T_5$ defects with respect to $T_3$ \cite{JNCS89,SCP91}.

The ideas of Pantelides  have then  been applied mainly in discussing the 
{\it geometrical} characterization of defects
\cite{InHoLee,tuttle96,Fedders87,Fedders92,tuttle98}.
Our aim is to discuss their soundness in terms of electronic properties,
by analyzing charge distributions and  atom-projected 
DOS obtained from  accurate {\it ab initio} calculations,
 also following some possible process of
formation  of $T_5$ defects and of evolution upon hydrogenation.

To this purpose,
we start from some selected samples generated by other
authors \cite{JNCS89,SCP91,BCCP91} using Car-Parrinello 
molecular dynamics (CPMD).  
These structures are a good starting point 
for this study, since they contain floating bonds; furthermore,
they reproduce quite well the
experimental pair correlation function and bond angle distribution
function using a reasonable number of atoms and hence they are
suitable for accurate ab-initio studies.
The small number of defects, which is however larger than in experiment,
 allows to easily single out effects associated to local features.
The configurations studied are 
cubic supercells of side $a=2\ a_0$, where $a_0$ is the
 equilibrium lattice parameter of c-Si. With respect to the original
configuration, where $a_0$ was fixed to the experimental value, we use the
theoretical equilibrium lattice parameter $a_0=10.17 a.u.$, 
which also corresponds ---in our calculations--- to the optimized
density of a-Si and a-Si:H. The starting configurations contain
respectively 64 Si atoms to describe a-Si \cite{JNCS89,SCP91} and
64 Si atoms plus 8 H atoms for a-Si:H \cite{JNCS89,BCCP91}.

We have studied both the mean configuration at room temperature and  
a snapshot of the CPMD run, in order to check for possible anomalies due to
statistical average, and also to single out the effect on the
electronic features of tiny structural variations. 
The CPMD configurations, aiming mainly at reproducing the structural 
properties, have been obtained using a kinetic energy cutoff $E_{cut}$=12 Ry
and the $\Gamma$ point only for Brillouin Zone (BZ) sampling
\cite{JNCS89,SCP91,BCCP91}.
Since  electronic structure studies require
better accuracy,  we improve  in our calculations the BZ sampling using
4 inequivalent special {\bf k} points for self-consistency and 
75  {\bf k} points for DOS.
These parameters have been chosen as
a reasonable compromise between accuracy and computational cost, after 
tests with  $E_{cut}$=16 Ry 
and with
the $\Gamma$ point  or   32 special {\bf k} points
for self-consistency.
We have used for Si the pseudopotential by Gonze {\it et al.} \cite{gonze} 
and for H a smoothed Coulomb potential.

The optimization of the a-Si and a-Si:H structures with the new computational
parameters 
is accompanied only by small structural rearrangements. 
The results for the structural and electronic properties
of the {\em mean relaxed} 
configurations  that we present here are essentially valid also for the others
(snapshot, unrelaxed).

The mean structural properties of these configurations are
similar to those discussed in refs. \cite{JNCS89,SCP91,BCCP91}.
We only report here that in a-Si the mean bond length is $d
\simeq 4.47$ a.u., quite similar to the
crystalline one which is 4.40 a.u.. 
The location of the first minimum of the radial distribution function 
defines geometrically the cutoff distance for the nearest neighbors (NN), 
which for Si-Si turns out to be  $R_{NN} = 5.08$ a.u. and $R_{NN} = 5.49$ a.u.
in a-Si and a-Si:H respectively.
In a-Si:H  each H is bound to one Si
atom with an average distance $d_H =$ 2.95 a.u..
With those values of $R_{NN}$, the resulting  average coordination 
number for Si in a-Si and a-Si:H is slightly larger than 4.
For the purpose of the present work a classification of coordination
defects based 
on a pure {\em geometrical} analysis is sufficient, but we point out that 
possible ambiguities in the bonding configuration can be resolved
only by an accurate analysis of the valence charge distribution and
of the ``electron localization function'' \cite{Torino,todo}.

\begin{figure}
\psfig{file=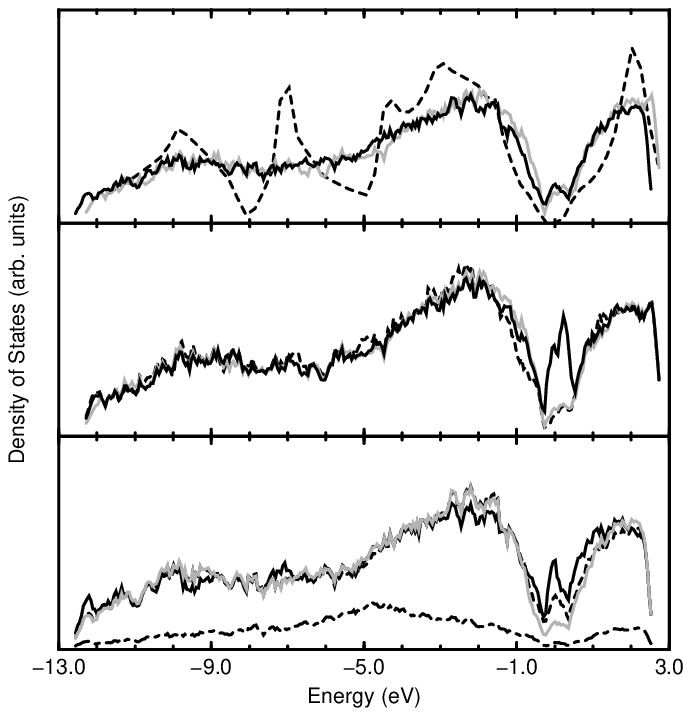}
\vspace{-7.4cm}
\hspace{7.5cm}
\psfig{file=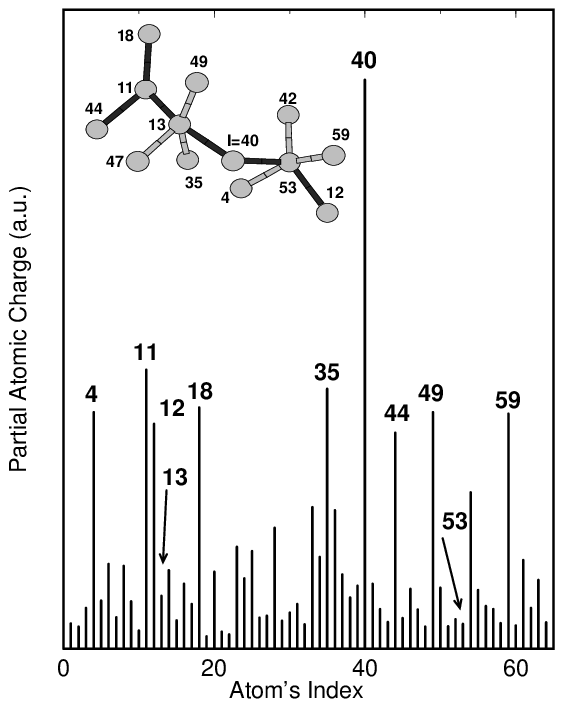}

\caption{Total and projected
DOS of c-Si, a-Si and a-Si:H.
DOS is normalized to unity and projected DOS to the number of valence
electrons for atomic type.
The energy scale is referred to the Fermi level.
Upper panel: total DOS
for a-Si (solid gray line), a-Si:H (solid black line), and
c-Si (dashed line).
Middle panel: atom-projected DOS for a-Si on $T_5$ atoms (dashed line),
$T_4$ atoms which are NNs of $T_5$ (solid black line), and the remaining $T_4$
atoms (solid gray line).
Lower panel: atom-projected partial DOS for a-Si:H on
$T_4$ atoms which are NNs of $T_5$ (solid black line),
$T_5$ atoms which are NNs of $T_5$ (dashed line),
the remaining $T_4$ and $T_5$ atoms  (solid gray line),
hydrogen atoms (dotted-dashed line).
}
\label{fig1}

\caption{Partial atomic charge
corresponding to
the occupied states in the gap
of a-Si. Almost all the
relevant contributions are associated to some NNs of the $T_5$
atoms, i.e. to atoms labeled 4, 11, 12, 35, 40, 49 and 59, which are NNs of
atom 13 or 53, or of both, as it is the case of atom
$I=40$. The peculiar local structure is sketched.
``Long'' bonds (see text) are in black.}
\label{fig2}
\end{figure}

Both our a-Si and a-Si:H configurations have a metallic
character, with coordination-defect induced  gap states
(the calculated total DOS is reported in the upper panel of 
fig.\,\ref{fig1}).
The precise definition of the energy gap is one of the unresolved
problem for an amorphous system.  We have chosen to
identify bottom  ($E_G^{bot}$) and top ($E_G^{top}$) edges of the gap 
with the absolute minima of the DOS below and above  the Fermi level $E_F$.
The occupied states in the gap  are the most relevant for our discussion,
 even if
their contribution to the charge density is small
(less than 1 \% for a-Si and 0.5\% for a-Si:H). 
Their atomic origin can be identified through the projection
on LCAO pseudo-wavefunctions.
The accuracy of the projection,
evaluated by means of the {\em spilling} parameter $S$ \cite{Sanchezportal96}
measuring the incompleteness of the LCAO basis
set, is satisfactory ($S$=0.03).

We focus now our attention on a-Si.
The configuration considered here contains two $T_5$
sites  close one to each other, with an
``interstitial'' ($I$) atom connecting them
(see fig.\,\ref{fig2}), and no ``true'' $T_3$ defects.
A charge density analysis helps in characterizing the different types of
bonds \cite{SCP91,Torino}. 
We  observe that $T_5$ sites are accompanied by a 
valence charge density depletion.
In particular,  those  $T_5$-$T_4$ bonds which are about 10\% longer
than the average value (``long'' bonds)
 and the bonds $T_5$-$I$ are characterized by a 
charge density distribution remarkably smaller 
(about 1/3) than the corresponding one
in the crystalline phase, or than the average in the amorphous phase.
These bonds are therefore weak and
are  the best candidates to break under a network distortion,
giving raise to a $T_4 + T_5$ $\rightarrow$
 $T_3+T_4$ transformation.

The middle panel of fig.\,\ref{fig1} shows the DOS projected
 on different classes of atoms:
$T_5$ atoms,  atoms which are NN of $T_5$, and the remaining $T_4$ atoms.
The peak of the DOS in the gap is  unambiguously associated to 
the NNs of $T_5$ sites.
In fig.\,\ref{fig2} we show the atomic projection of the charge density of the
 gap states, obtained again from the LCAO coefficients. It is even more clear
that the gap states are not located on $T_5$'s themselves but are delocalized
on their NNs, and in particular on those
  connected  by longer and
weaker bonds. 
  Evidence of a midgap state with these characteristics is also reported in 
ref. \cite{tuttle98}.
We argue that 
previous works ascribed the origin of gap states uniquely \cite{InHoLee}
or mainly \cite{fedders88,bisw89,tuttle96,Fedders92,drabold90,hm92} to $T_3$,
simply because they  disregarded an accurate analysis of NN environments.

From inspection of fig.\,\ref{fig2}
we observe the large contribution to the DOS in the gap coming from
the atom $I$ which is NN of two $T_5$'s. We also observe
 that also some normally coordinated sites (labelled 18 and 44), 
characterized by having 
``long'' bonds, partially contribute to gap states, in agreement with
 similar findings reported
 in refs. \cite{bisw89,tuttle96,drabold90}.

The a-Si:H \cite{BCCP91}  structure  studied here is more complex.
Well defined $T_5$ sites are present and are the predominant kind of defect
(we do not address in this paper the discussion of other more complicated
defects which could require a better characterization beyond
the pure geometrical analysis \cite{Torino,todo}).
Overcoordination arises from 
five Si atoms, or from four Si atoms and one hydrogen as NN.
The main features of the electronic structure 
are not  very different from those of a-Si.
As we found in a-Si, those atoms which 
are NNs of some $T_5$ sites give a large contribution to the DOS
in the gap (fig.\,\ref{fig1}, lower panel).

In order  to get further
insight onto the characterization of the electronic properties of $T_5$
defects and to recover the ideas of Pantelides, 
we want to follow in terms of electronic states some possible
processes of $(a)$ $T_3\rightarrow T_5$ transformation 
and $(b)$ annihilation of $T_5$ by H.
To this purpose, we follow two ``computational experiments'' starting
 from the CPMD configurations described before.

\begin{figure}
\psfig{file=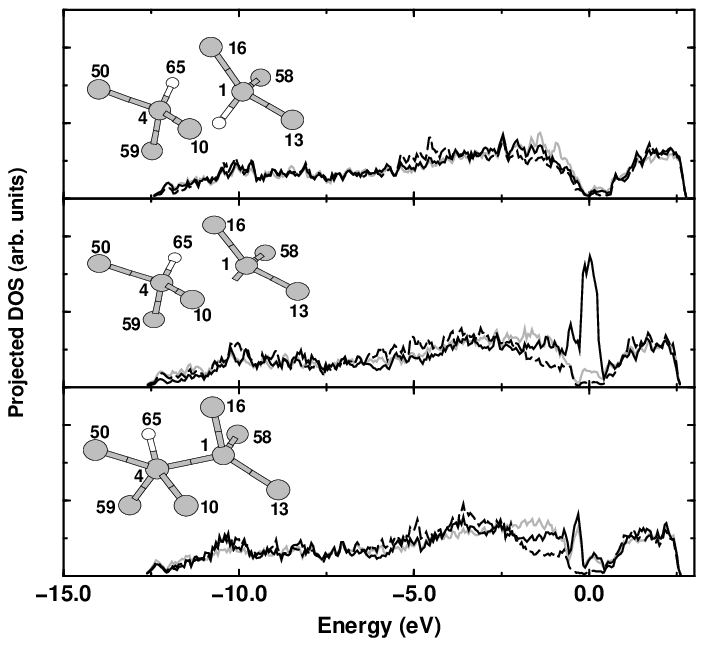}
\vspace{-6.7cm}
~\hspace{7.cm}
\psfig{file=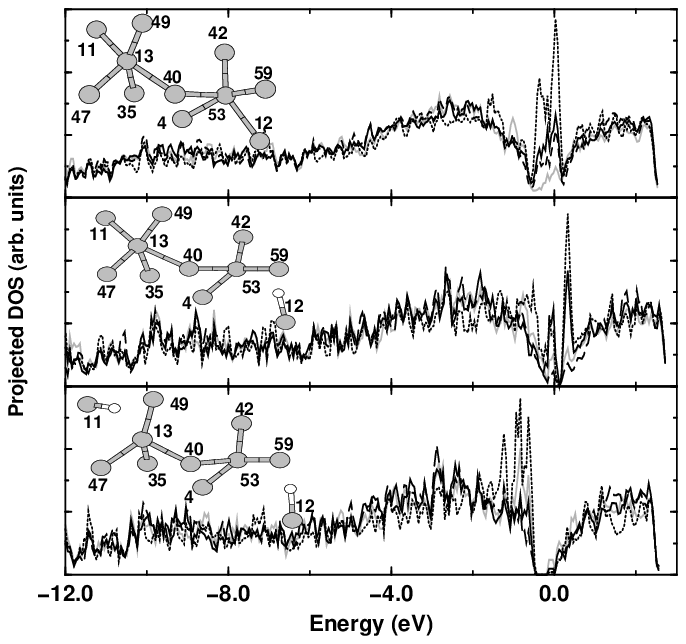}

\caption{Projected DOS of a cluster 
of Si (gray) and H (white) atoms selected
from a-Si:H, evolving from a $T_4+T_4$ (upper panel) to a $T_4+T_3$
configuration (central panel) after removal of one H atom, 
and finally to a $T_5+T_4$ configuration (lower panel) after optimization of
atomic positions. Solid black line is for atom 1, dashed line for atoms 4,
13, 16, 58 (whose contribution to the DOS remains essentially unchanged),
grey line for atoms 59, 10, 50, 65, which are  NNs of atom 4.
}
\label{T3toT5}

\caption{Projected DOS of the cluster of atoms
selected from a-Si initially containing the two $T_5$ defects (upper panel),
and after the insertion of a first H nearby the $T_5$ labelled 53
 (central panel) and a second one nearby the $T_5$ labelled 13 (lower panel),
after optimization of atomic positions. 
Dotted line is for atom 40, 
solid black line is for the remaining NNs of atom 13, 
dashed line for  NNs of atom 53,
thin gray line for atoms 13 and 53, 
whose contribution to the DOS in the gap is negligible and
remains essentially  unchanged in the evolution of the structure.
}
\label{annihilation}
\end{figure}

In the first one, we start from a-Si:H sample. 
We focus our attention on a normally coordinated cluster of atoms  
(fig.\,\ref{T3toT5}, upper panel) containing two $T_4$
Si (labelled 1 and 4)
close one to each other ---but not directly connected---
and their NNs which include two H atoms. 
This structure $T_4+T_4$ 
has an almost negligible DOS in the gap.
We first create artificially a $T_3$ site by removing 
one  H atom and leaving
all the other atoms fixed, so that the structure is now $T_4+T_3$
(fig.\,\ref{T3toT5}, central panel). 
This configuration corresponds  to a huge DOS in the gap,
almost entirely originating from the dangling bond. After relaxation of the
atomic positions the distance between atoms 1 and 4 is reduced from
5.63 a.u. to 4.48 a.u. and a new bond is formed, so that the structure  
is now  $T_5+T_4$ 
(fig.\,\ref{T3toT5}, lower panel).  
This process (removal of one H followed by bond
reconstruction) is similar to that studied in ref. \cite{tuttle98},
with the difference that the other H atom in our configuration is replaced
there by one Si atom. We observe that in the final configuration ($T_5+T_4$)
there are still midgap states, 
but now they are delocalized on the NNs of the atom 4 and
pushed towards lower energies.
Our findings support  the picture of conjugated $T_3$ and $T_5$ 
defects. Moreover,  the spontaneous evolution from $T_4+T_3$ to $T_5+T_4$,
which is in agreement with the findings of ref.~\cite{Fedders87},
confirms --although in a particular case--- that
 the $T_5$'s can be energetically 
favoured with respect to the $T_3$'s, as indicated in ref. \cite{KT88}.

The inverse process, that may occur in real systems
as proposed in ref. \cite{Pantelides87},
indicates a possible mechanism of annihilation of $T_5$ defects by H.
But this can be better studied following in terms of electronic states
our second computational experiment.
We start from the a-Si configuration and we focus our
attention on the cluster with the two $T_5$ sites (fig.\,4, upper panel). 
We add one H atom nearby a $T_5$. Relaxing the whole  structure, a $T_4-T_5$
``long'' bond breaks, so that a new
Si-H bond is formed and the $T_5$ becomes a normal $T_4$ site
 (fig.\,4, middle panel). This is accompanied by a lowering of the DOS in the
gap.
Adding another H atom, also the second $T_5$ is annihilated and the
final structure has a vanishing DOS in the gap (fig.\,4, lower panel).

A variety of other structural models  have been proposed in the literature
for  a-Si and a-Si:H, describing even better some  global properties,
but we are confident that our findings  concerning local environments 
are not much affected by the  particular choice of the structural model.
Infact preliminar results of another  work in progress \cite{todo}
using a larger sample of a-Si:H
generated by L. Colombo using tight-binding molecular dynamics \cite{colombo}
confirm the validity of the present findings.

In conclusion, we 
have presented the results of  accurate {\it ab initio} self-consistent
pseudopotential calculations of the electronic properties of $T_5$ defects 
in a-Si and a-Si:H starting from some configurations generated via
{\it ab initio} molecular dynamics, and we have followed in terms of
electronic density of states some possible processes of their formation and
annihilation by H.  
Our calculations clearly indicate that
$i)$ $T_5$ sites give a {\em fundamental} contribution to gap states through
their NNs (fig. 1, 2);
$ii)$ in some circumstances the evolution from $T_3$ to $T_5$ is favoured,
and this is accompanied by a delocalization of the corresponding 
electronic states with a shift towards lower energies (fig. 3);
$iii)$ the DOS in the gap can be
reduced by the interaction of H with the  $T_5$'s (fig. 4).
These results  support Pantelides' proposals, and suggest that 
$T_3$ and $T_5$ in a-Si  not only are ``structurally''
conjugated, but they also play a similar role in originating gap states,
which can be in both cases passivated by hydrogen.
This pictures reconcile those experimental findings
that a description  uniquely emphasizing the dangling bonds would not explain.

\stars
This work has been done within the ``Iniziativa Trasversale di Calcolo
Parallelo'' of INFM using the parallel version of the PWSCF 
(Plane-Wave Self-Consistent Field) code.
One of us (SdG) acknowledges support from the MURST
within the initiative ``Progetti di ricerca di rilevante interesse
nazionale''.
We would like to thanks {\sc L.~Colombo} and
{\sc G.~L.~Chiarotti} for useful discussions and suggestions.


%
\end{document}